\documentclass[conference]{IEEEtran}
\pdfoutput=1
\usepackage{amsmath}
\usepackage{graphicx}
\usepackage{booktabs}
\usepackage{adjustbox}
\usepackage{cite}
\usepackage{float} 
\usepackage{tabularx} 
\usepackage{multirow}
\usepackage{hyperref}
\usepackage{afterpage}
\usepackage{caption}
\usepackage{subcaption}
\usepackage{verbatim}
\usepackage{ragged2e}
\usepackage{lipsum,framed}

\usepackage{xcolor}
\usepackage{graphicx}
\definecolor{cl1}{rgb}{0.8,0.9,1}

\begin{document}

\title{MERCI: Multimodal Emotional and peRsonal Conversational Interactions Dataset}

\author{Mohammed Althubyani$^1$, Zhijin Meng$^2$, Shengyuan Xie$^2$,  Cha Seung$^2$, Imran Razzak$^{3,2}$, \\ Eduardo B. Sandoval$^4$,$^2$, Baki Kocaballi$^1$, Francisco Cruz$^{2,5}$\\
$^1$School of Computer Science, University of Technology, Sydney, Australia \\
$^2$School of Computer Science and Engineering, University of New South Wales, Sydney, Australia \\
$^3$Mohamed bin Zayed University of Artificial Intelligence, Abu Dhabi, UAE \\
$^4$School of Art and Design, University of Technology, Sydney, Australia \\
$^5$Escuela de Ingenier\'ia, Universidad Central de Chile, Santiago, Chile
}
\maketitle

\maketitle
\begin{abstract}
The integration of conversational agents into our daily lives has become increasingly common, yet many of these agents cannot engage in deep interactions with humans. Despite this, there is a noticeable shortage of datasets that capture multimodal information from human-robot interaction dialogues. To add a systemic gap, we have recorded a novel multimodal dataset that encompasses rich embodied interaction data (MERCI). The process involved asking participants to complete a questionnaire and gathering their profiles on ten topics, such as hobbies and favourite music. Subsequently, we initiated conversations between the robot and the participants, leveraging GPT-4 to generate contextually appropriate responses based on the participant’s profile and emotional state, as determined by facial expression recognition and sentiment analysis. Automatic and user evaluations were conducted to assess the overall quality of the collected data. The results of both evaluations indicated a high level of naturalness, engagement, fluency, consistency, and relevance in the conversation, as well as the robot's ability to provide empathetic responses. It is worth noting that the dataset is derived from genuine interactions with the robot, involving participants who provided personal information and conveyed actual emotions.
The code and the dataset is publicly available at [ANONYMOUS]. 

\end{abstract}

\section{Introduction}

Effective communication incorporating verbal and non-verbal cues is crucial for seamless human-robot interaction \cite{shimoda2022role}. However, much of the research in this area has concentrated on developing domain-specific conversational systems, with limited attention to open-domain dialogue for social robots. As AI systems become more integrated into everyday life, their ability to engage users in meaningful and personalized ways is critical.

Traditional conversational agents, often rule-based, operate by following a set of pre-programmed commands and keywords. This rigid approach limits their ability to adapt to diverse and unpredictable interactions, reducing the scope of engagement with human users \cite{cherakara2023furchat}.
In contrast, the recent advancements in large language models (LLMs) have ushered in a transformative shift in natural language processing (NLP). Recently, models like GPT-4 have made significant breakthroughs, particularly in tasks like question answering, text summarization, and dialogue generation, showcasing a remarkable ability to understand and generate human-like text \cite{alawida2023comprehensive}. These LLMs can handle a broader range of topics and conversational nuances than traditional rule-based systems. 
The primary obstacle to implementing LLM is developing systems that can smoothly transition from closed-domain interactions—focused on specific, predefined topics—to open-domain conversations, where the subject matter is unpredictable and wide-ranging \cite{kong2024opentab}. 
It is crucial to consider the embodiment of these systems, as integrating LLMs into physical agents like robots introduces further complexity. This requires sophisticated coordination between linguistic capabilities and physical actions for seamless human-robot interactions.

 By creating a dataset that incorporates personality and emotional nuances, this work aims to contribute to the advancement of HRI research, providing researchers and developers with the tools to train AI systems capable of delivering more meaningful, personalized experiences. 
 We present a large-scale multi-modal robotics conversational dataset collected from 30 participants and develop a conversational system named PERCY, implemented on the ARI robot platform. \cite{pal_robotics_ari}. PERCY is designed for personalized and emotionally aware Human-Robot Interaction (HRI). Unlike traditional systems that facilitate basic conversations, PERCY can interpret users' emotional states and respond empathetically, using real-time facial expressions, vocal tone, and sentiment analysis. PERCY enhances the conversational depth and generates a novel dataset that captures two critical aspects of human-machine interaction: personal information and emotional cues. This dataset will be a valuable resource for training AI models and producing a more personalized and emotionally engaging dialogue. The key contributions of this research are as follows:


\begin{itemize} 
\item  Introduce robotic conversational multimodal dataset MERCI collected from 30 subjects, derived from real interactions with participants, incorporating their personal information and emotions. 

\item Develop a conversational framework PERCY using  a personal emotional robot capable of identifying user emotions and providing empathetic responses through real-time facial expression analysis, sentiment analysis, and personalized information. 

\item Provision of a comprehensive analysis and insights into the efficacy of integrating personal information and emotional cues to enhance the personalization and empathy of conversational systems.
\end{itemize}
 This research paves the way for future advancements in emotionally intelligent AI-human interactions.

\section{Related Work}
\subsection{Conversational System}
Conversational systems, often referred to as dialogue systems or chatbots, have become an integral part of daily human-computer interactions \cite{mctear2016conversational}. These systems aim to simulate human conversation by understanding and responding to user inputs through text and speech interactions \cite{syvanen2020conversational}. Despite significant advancements, many conversational agents cannot still provide truly personalized and engaging interactions. While early systems like ELIZA \cite{bradevsko2012survey} laid the groundwork for human-computer dialogue, modern AI-based systems have taken these capabilities further by offering more sophisticated language processing and contextual understanding \cite{vaswani2017attention}. Nevertheless, there is still a growing need to enhance personalization in these systems to make interactions more human-like and empathetic. Areas that require further research are:

\subsubsection{Personalization of The Dialogue System}
Personalized dialogue systems are critical for creating meaningful and engaging conversations, particularly in human-robot interaction (HRI). In traditional systems, most conversations are based on generic datasets, which limits their ability to tailor responses to individual users. As a result, these interactions often feel unnatural and impersonal \cite{boyd2020large}. Personalization in conversational systems involves adjusting responses based on a user’s unique preferences, interests, and emotional state. This process requires the system to retain and apply contextual information, such as the user's past interactions, preferences, and personality traits in real-time. Missing that, conversational agents risk becoming disengaged and unrelatable, ultimately failing to foster a genuine connection with users \cite{liu2020you}.

\subsubsection{Incorporating Empathy into Personalized Dialogue Systems}
A crucial aspect of personalization in human-robot interaction is the ability to detect and respond to human emotions. Emotional awareness enables robots to understand the user's feelings and adjust their responses accordingly, making interactions more human-like and supportive. For dialogue systems to be genuinely engaging, they must recognise and respond to human emotions, going beyond simple sentiment analysis \cite{zhu2021topic}.
This is particularly important in human-robot interaction, where robots are expected to perform tasks beyond basic communication, such as providing companionship, offering customer service, or even assisting in healthcare settings \cite{sirithunge2021evaluation}.

Recent advances in emotional detection, such as facial expression analysis, voice modulation, and sentiment analysis, allow robots to respond empathetically, significantly improving user engagement \cite{park2022empathy}. Empathy allows robots to understand a user's feelings and reflect those emotions in their responses, fostering deeper emotional connections. Additionally, studies have shown that users respond more positively to robots that demonstrate emotional awareness and adapt their behaviour accordingly \cite{mishra2023real}. Integrating empathy into dialogue systems makes the interaction feel more natural and improves user satisfaction by making the system appear more understanding and supportive.

\subsection{Dialogue Dataset}
There are two kinds of dialogue systems: domain-specified and open-domain dialogue. The most traditional datasets are domain-specified \cite{chen2019jddc}, which are of small scales and oriented to complete specific tasks like flight booking, hotel reservation, and customer services.

Recent datasets are more open-domain, which is much more challenging to develop due to the open-ended goal. More specifically, it is difficult to calculate the optimization of open-domain dialog because there are many dialog skills to enhance the engagement that require a deep understanding of dialog context and the user’s emotion to select the most appropriate dialog skill in real-time to generate the most optimal responses to align with the user’s emotional state and maintain a consistent personality \cite{huang2020challenges}.

For instance, Daily Dialog \cite{li2017dailydialog} is a multi-turn dialogue dataset, that contains 13,118 dialogues crawled from various websites that serve English learners to practice English dialog in daily life.  It has statistics of average speaker turns per dialogue as 7.9, average tokens per dialogue as 114.7, and average tokens per utterance as 14.6.  This dataset is rich in emotion due to the two main reasons why people communicate in daily life, namely exchanging information and enhancing social bonding, where emotional expression is needed. Although Daily Dialog is full of emotional states, it lacks multi-model information like video and audio.

In this regard, MELD Dataset \cite{poria2018meld} fixes the weakness of Daily Dialog and it is the most relevant dataset to the target of this research project. MELD developed a dataset of more than 1,400 dialogues and 13,000 utterances with emotional annotations but containing additional multi-model information including visual and audio counterparts from the TV series ‘Friends’. It contains several emotional states and multi-model information since it is extracted from the TV series. However, it is crucial to recognize that it may be susceptible to the biases of TV series screenwriters, potentially introducing artificial elements into the dialogues, thereby hindering its practical applicability in real-life scenarios. 
\subsection{Contribution of this work}
Despite the progress in personalization and emotion detection, significant challenges remain. Many systems still struggle with maintaining consistency in their personalized responses across multiple interactions, and emotional cues are often not fully integrated into the conversation flow \cite{kim2020will}. Moreover, creating a dynamic and adaptive system that can handle the complexity of human emotions while maintaining coherence and relevance throughout the conversation is a challenging task. Our approach to overcoming these limitations is the integration of multi-modal data that includes both personal information and emotional states. By combining these elements, robots can generate responses that are not only contextually relevant but also emotionally aligned with the user's needs and preferences \cite{ma2020survey}. This approach creates a deeper connection between the user and the robot, fostering more meaningful interactions.

This research project conducted data acquisition from interactions between humans and the ARI robot. The research procedure is justified by the need to collect user profile information from individuals engaged in natural conversations, thereby creating a realistic dataset representative of how people interact. Furthermore, employing a robot to gather this data is justified as it allows for exploring human-robot interactions to enhance such relationships while enabling us to collect the data in a fully controlled environment. This approach will ultimately enhance the user’s experiences when interacting with AI systems, as these systems can be more adaptive to the individual’s preferences and emotional states. A key advantage of this dataset is to provide high-quality data suitable for various application domains, including AI, machine learning, natural language processing, psychology, and human-computer interaction due to its nature of being an open-domain dataset.

\section{PERCY Design}
\begin{figure}[!htb]
    \centering
    \includegraphics[width=\linewidth]{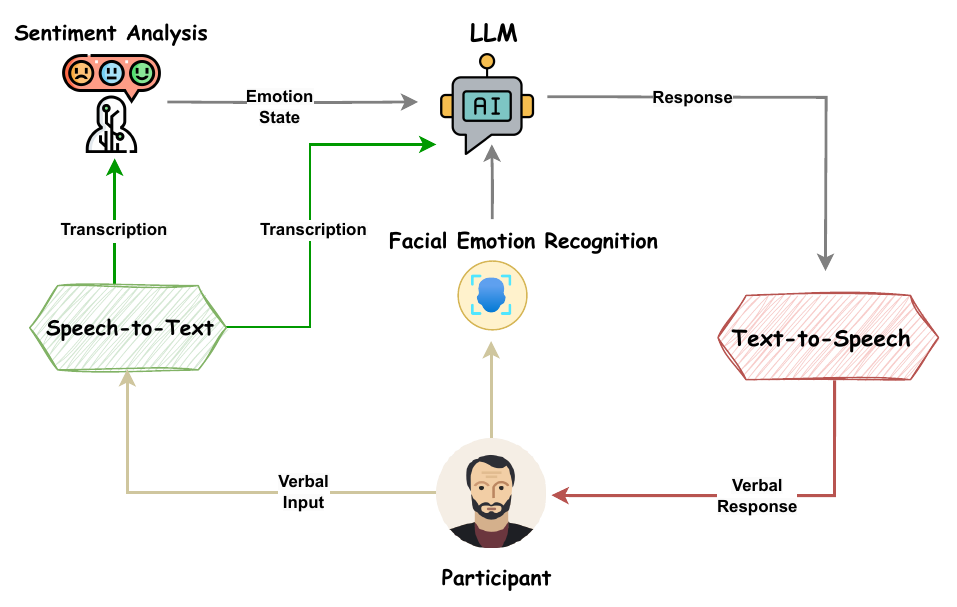} 
    \caption{PERCY analyzes users' emotional state through real-time facial emotion recognizer and sentiment analysis. Appropriate responses will be generated based on the users' emotional state and finally, PERCY will speak to the response back to the user. }
    \label{fig:ros_design}
\end{figure}

PERCY consists of five modules: (I) speech-to-text module, (II) large language model-based module (LLM–GPT), (III) facial emotion recognition module, (IV) sentiment analysis module, and (V) text-to-speech module.  Please see figure \ref{fig:ros_design}.
The interactions among the above modules are conducted through 3 ROS1 servers, namely the main server, the emotion server, and the GPT server. Module (I), (IV), and (V) are integrated into the main server. When participants speak a sentence, the speech-to-text module will pass the transcription to both the sentiment analysis and large language model-based modules. Meanwhile, the facial emotion recognition module will analyze participants' emotional state from the live video stream received from the depth camera on the ARI robot's head. With the transcription and the emotional states (from sentiment analysis and facial emotion recognition) as input, the GPT server will generate the response. Finally, the generated response is spoken to the user through the text-to-speech module. Following is the description of each module in PERCY.


\subsection{Speech-to-Text Module}

To convert participants' speech into text for input into the GPT-based system, the PERCY framework utilizes the OpenAI Whisper API \cite{openai_whisper} as its speech-to-text module. Instead of publishing the accuracy, OpenAI declared that whisper outperforms the previous models. Although it's sensitive to the users' accents, whisper turned out to perform well in most cases in this project. 

\subsection{Large Language Model-based Dialogue Management}
The speech-to-text module sends the participant's spoken words as input to the ChatGPT-4 API. Setting an appropriate prompt is crucial, as it directly influences the generation of tailored, empathetic responses in human-robot interactions. The prompt provides PERCY with clear instructions on how to engage in meaningful, natural, and coherent conversations. It emphasizes empathy, fluency, and relevance in every interaction, ensuring that the robot's responses align with the user’s profile and emotional state. By incorporating real-time emotion detection and user-specific information, the prompt enables PERCY to deliver contextually appropriate questions and responses, fostering deeper engagement and a more human-like conversational experience. This approach allows PERCY to create seamless, adaptive interactions that respond not only to users’ inputs but also to their emotions, making the conversation more personalized, empathetic, and engaging (see Additional details in Appendix A).

\subsection{Sentiment Analysis Module}
The NLTK package \cite{nltk} is used as the sentiment analysis module. With the use of VADER, NTLK analyzes the output of the speech-to-text module and outputs the participants' emotional status to the GPT-based system. Then the GPT-based system could consider the sentiment analysis emotional status in generating the response.

\subsection{Facial Emotion Recognition Module}
A real-time facial emotion module is integrated into PERCY. It receives the real-time video data stream from the depth camera mounted on the ARI robot's head. This system further provides the participant's emotional status (happy, sad, angry, confused, fearful, disgusted, neutral) to the GPT-based system, enabling the robot to dynamically adjust its responses based on the user's emotions, creating a more empathetic and engaging interaction with the user.
The backbone network used in this system is based on MobileNetV2, which efficiently extracts multi-scale feature maps from the input images. These feature maps represent the different levels of detail in the image, enabling the model to detect subtle emotional cues in the user's facial expressions. The network is structured to produce feature maps from multiple convolutional layers, allowing for robust emotion classification at various scales.

\subsection{Text-to-Speech Module}
The built-in ARI robot rostopic \cite{ari_sdk} is used as the text-to-speech module, enabling the robot to speak the responses generated by the GPT-based system to the participant.

Further details related to the implementation and Pipeline Data Testing will published in following paper shortly. 

\section{Dataset Collection}

\subsection{Ethical Approval}
 The research team has received full ethical approval from the Human Research Ethics Committee (HREC) at the University of XXXX (UXXX).
 Approval number is XXXXXXXX.

\subsection{Experiment Procedures}
Data collection was conducted by recruiting participants to complete a three-step process: an initial questionnaire, a follow-up conversation with the ARI robot, and a feedback questionnaire.

\subsubsection{Initial Questionnaire}

The initial questionnaire collects non-sensitive personal data to build a detailed user profile that enhances subsequent human-robot interactions. By combining multiple-choice and open-ended questions about the user's personality, interests, and preferences, the questionnaire gathers essential information to customize the interaction. As highlighted by Gasteiger et al. \cite{gasteiger2023factors}, successful human-robot interaction hinges on the robot’s awareness of various human factors, such as user interests, preferences, personality, attitudes, educational background, and past experiences. This gathered information forms the basis for future conversations, allowing the robot to deliver more personalized and engaging responses.

\subsubsection{Follow-up Conversation}
\begin{figure}[ht]
    \centering
    \includegraphics[width=0.6\linewidth]{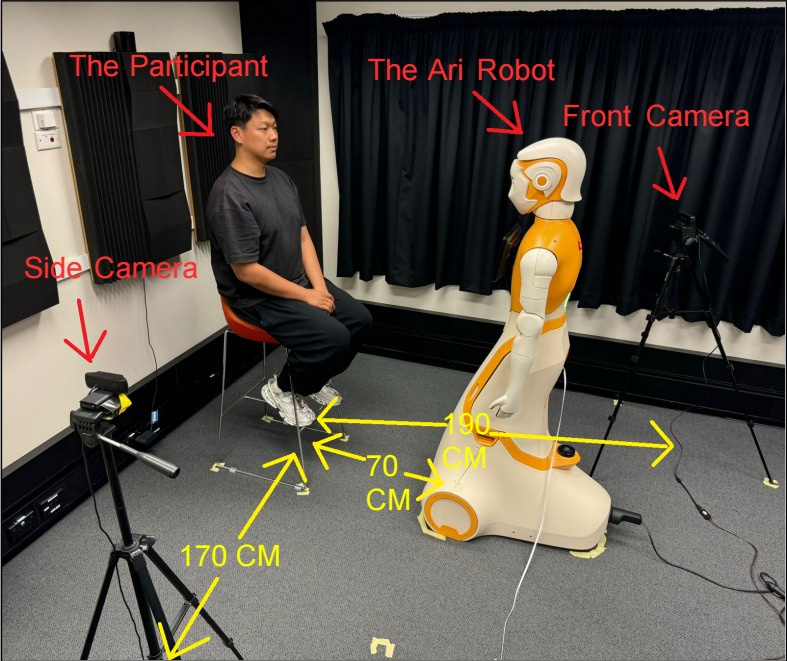} 
    \caption{In addition to the participant and the robot, two external cameras are positioned to record the experiment from the front and the side, providing a comprehensive overview of the conversation.}
    \label{fig:experiment_setting}
\end{figure}

\begin{figure}[htbp]
    \centering
    \begin{subfigure}{0.2\textwidth}
        \centering
        \includegraphics[width=\textwidth]{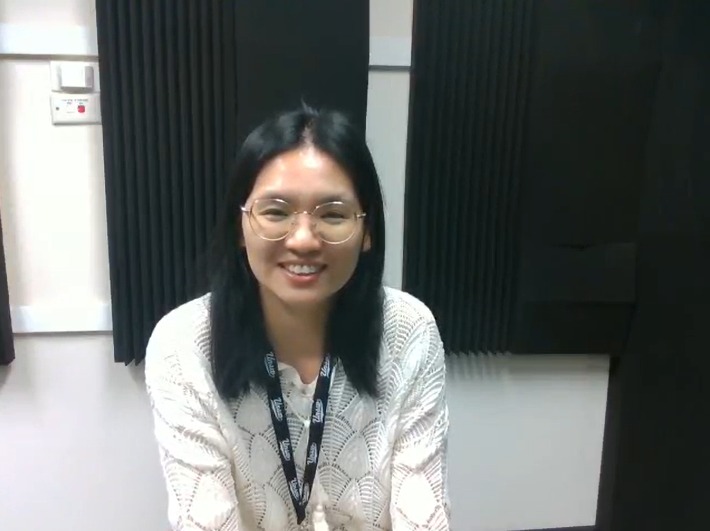}
        \caption{Robot Head Camera Video Channel}
        \label{fig:image1}
    \end{subfigure}
    \hfill
    \begin{subfigure}{0.2\textwidth}
        \centering
        \rotatebox{180}{\includegraphics[width=\textwidth]{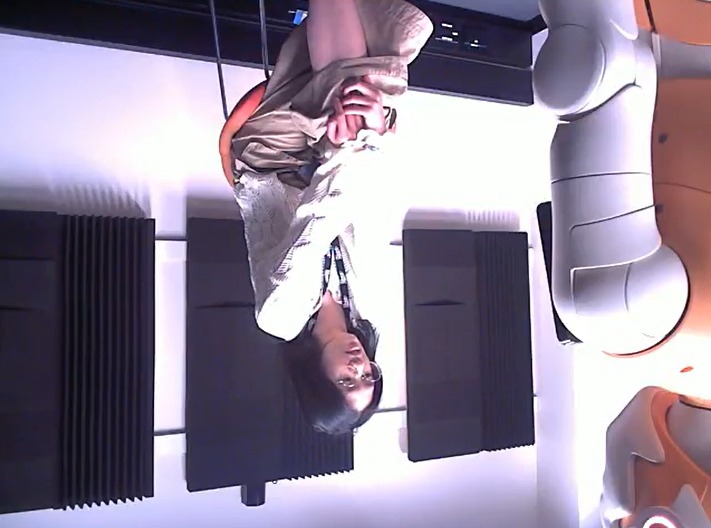}}
        \caption{Front Camera Video Chanel}
        \label{fig:image2}
    \end{subfigure}
    \hfill
    \begin{subfigure}{0.2\textwidth}
        \centering
        \includegraphics[width=\textwidth]{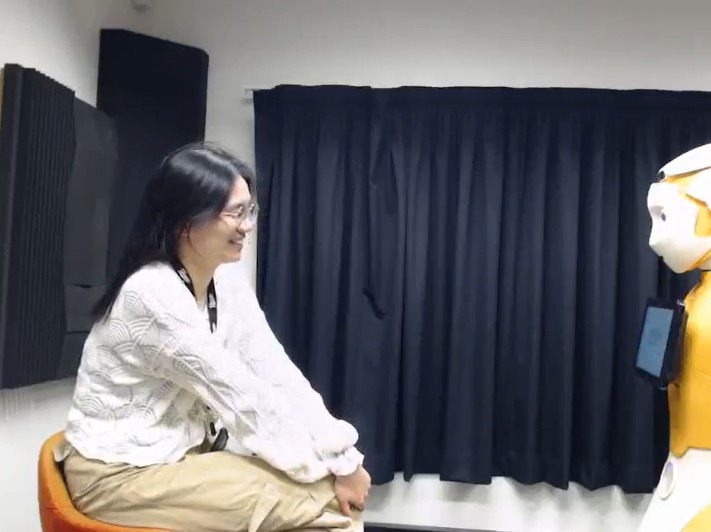}
        \caption{Side Camera Video Chanel}
        \label{fig:image3}
    \end{subfigure}
    \caption{The stream data from the ARI robot head camera (a), the front camera (b) and the side camera (c)}
    \label{fig:video_channels}
\end{figure}

The ARI Robot then uses PERCY to conduct the conversation with the participants, asking follow-up questions based on the personal data collected.  
During these interactions, participants are encouraged to express their personalities, preferences, and emotions naturally. 
The conversation setup includes the participant, the ARI robot, and two external cameras positioned at the side and front, shown in figure \ref{fig:experiment_setting}. 
There are three video channels: the ARI robot's head camera and the two external cameras, the stream data from these channels are shown in figure \ref{fig:video_channels}. Additionally, there is only one audio channel from the ARI robot microphone.

\subsubsection{Feedback Questionnaire}
Finally, participants are asked to complete the feedback questionnaire to collect their general impressions of the interaction with the robot. This will measure their overall experience in areas such as comfort, engagement, and the naturalness of the interaction.
The feedback questionnaire includes sections dedicated to measuring the NASA Task Load Index (NASA TLX) and Negative Attitudes Towards Robots (NARS), standard metrics used in Human-Robot Interaction. 
The NASA TLX assesses the user's workload during system interaction, while the NARS measures the user's negative attitudes towards robots.

\subsubsection{Example}
We designed a persona-based conversational robot. This implementation involves designing a system that can understand emotions and interact with users compassionately, clearly, and supportive. We named our robot Noori.  Noori, is a persona-based conversational robot that embodies a caring, patient, and calm personality. She avoids overwhelming users with too much information and engages them in familiar, comforting conversations. With a soothing voice and warm demeanor, Noori provides emotional support by detecting and interpreting users' emotions from their tone, word choices, and mood, adjusting her responses accordingly.


This is an example between our personalized conversational robot "Noori", and a hypothetical user "Sam" which highlights the emotional intelligence and personal engagement of the system:

\fcolorbox{red}{cl1}{
\begin{minipage}{22em}
\noindent \textbf{Noori:} Hi Sam! (emotion: anxious) Are you feeling a bit tired today? \\
\noindent  \textbf{Sam:} Noori.., I don’t know… I’ve just been feeling a little down lately. \\ 
\noindent  \textbf{Noori:} (emotion: sad) I’m really sorry to hear that Sam. It sounds like things have been tough for you. Do you want to talk about what’s been on your mind? Sometimes sharing helps. \\
\noindent  \textbf{Sam:} I just feel like everything’s too much. Work, life, everything... I can’t seem to catch a break. \\ 
\noindent  \textbf{Noori:} (emotion: netural) That sounds overwhelming. It’s completely understandable to feel that way when life gets busy. Would it help to talk through one of the things that's stressing you out right now? \\
\noindent  \textbf{Sam:} I guess work has been really tough. I’m struggling to keep up with all my projects. \\
\noindent  \textbf{Noori:} (emotion: netural) That sounds like a lot of pressure. Maybe we can break it down together.  \\
\noindent \textbf{Sam:} It’s this new presentation for a big client… I’m really nervous about it.\\

\end{minipage}

}

\section{Results}

In this section, we describe the gathered dataset and its automatic and user evaluations, as well as the evaluation criteria. 

\subsection{Participants}
\begin{figure}[ht]
    \centering
    \includegraphics[width=\linewidth]{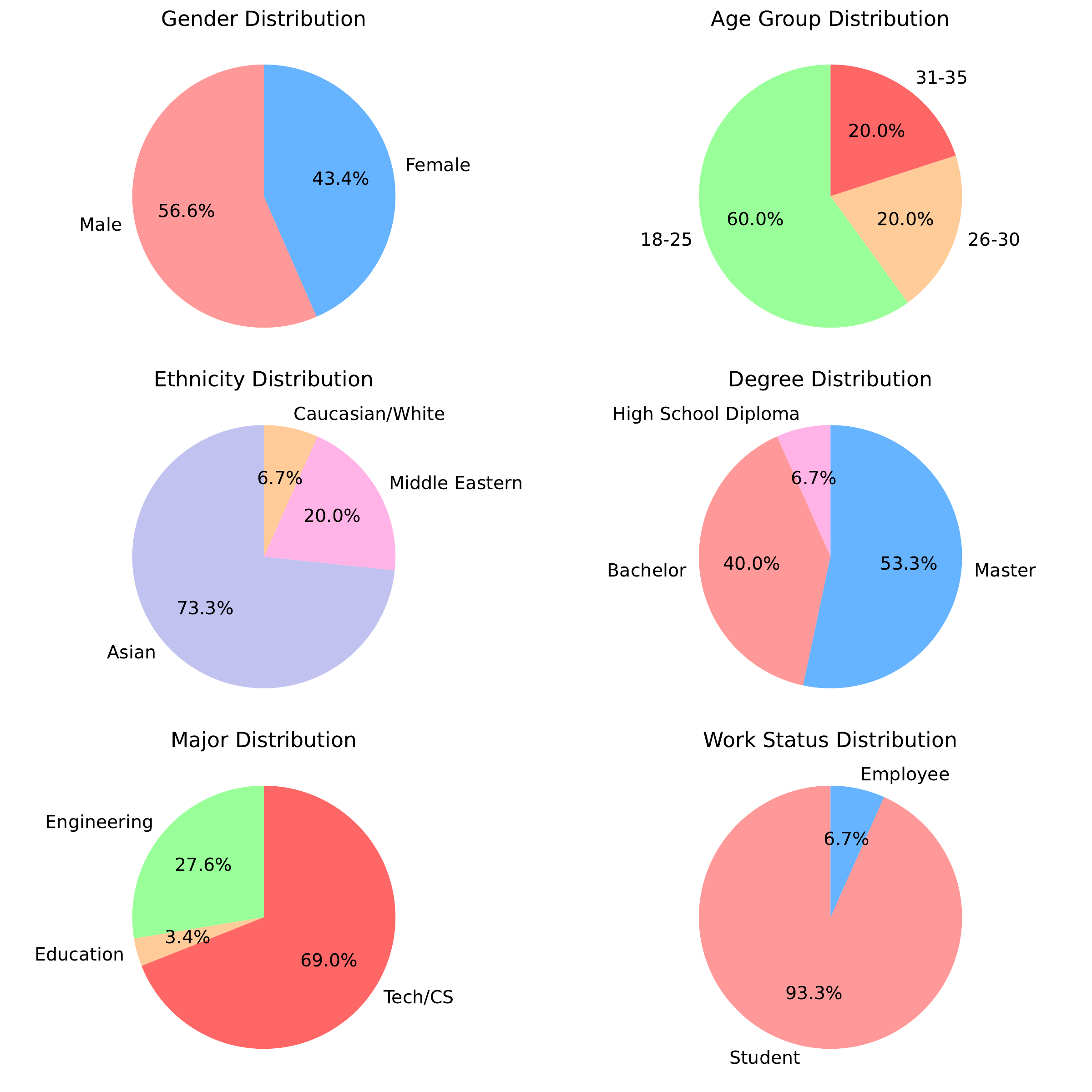} 
    \caption{The participant pool is diverse in terms of age, cultural background, and educational background.  .}
    \label{fig:participants}
\end{figure}

A total of 30 participants were recruited for this project, primarily from the XXXXXX campus. The participant pool is diverse, with balanced representation in terms of gender and age. As shown in Figure \ref{fig:participants}, the male participants represent 56\% (17 males) of the total number of the current participants, leaving 43\% for females. With ages ranging from 18-35 and different ethnicities including Asian, Middle Eastern, and Caucasian or White.  

\subsection{The Dataset}
The data collected contains recorded videos and audio of the entire participants' robot conversations, with an average of 20.27 minutes of conversation per participant. The longest recording time is 37.39 minutes while the shortest is 15.07 minutes. Each conversation consists of 62 utterances, making a total of 1860 utterances. 

\begin{figure}[ht]
    \centering
    \includegraphics[width=\linewidth]{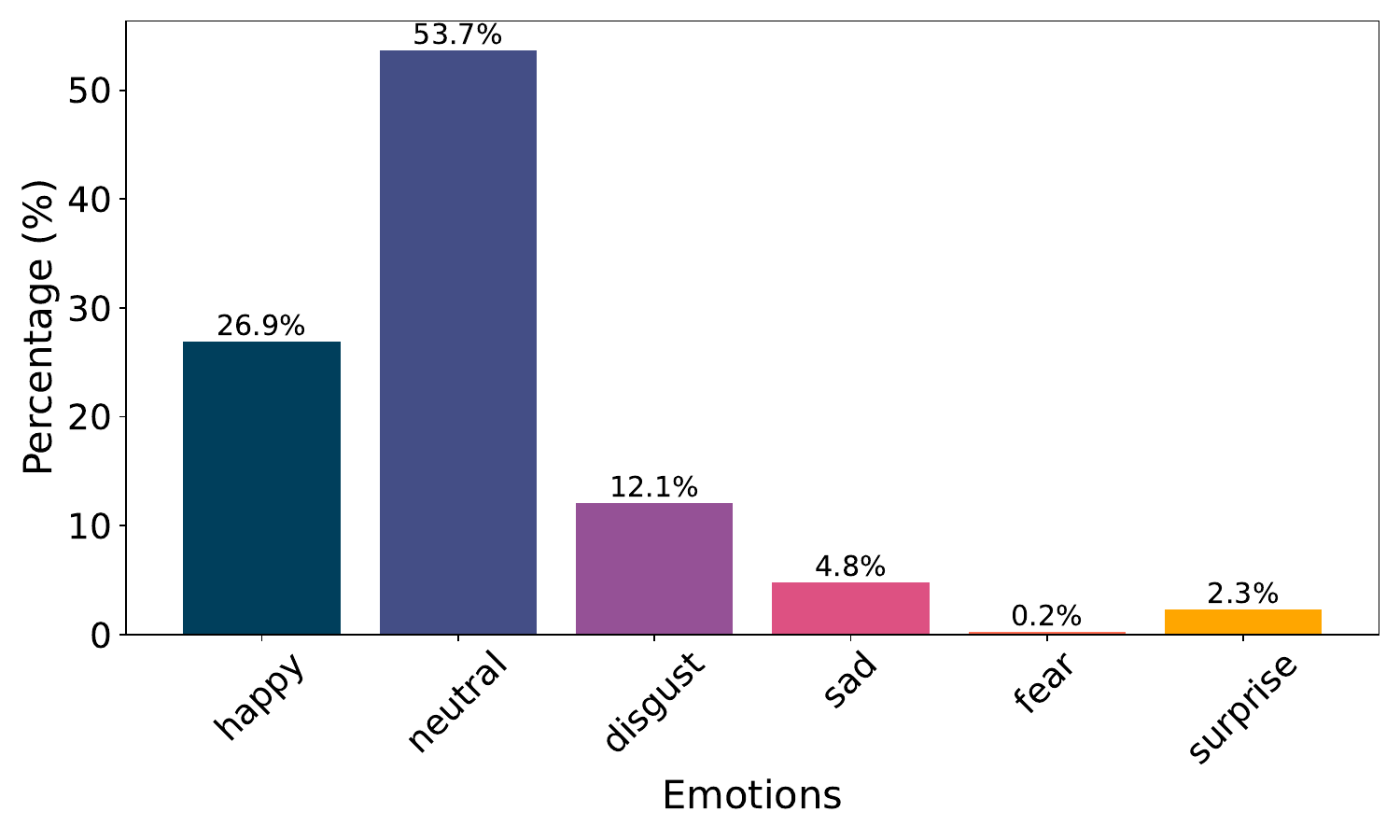} 
    \caption{Emotions distribution across all the conversations.}
    \label{fig:emotins_dis}
\end{figure}

This dataset contains a total of 900 emotions. The emotions distribution as illustrated in figure \ref{fig:emotins_dis} reveals that neutral emotions appeared the most frequently, with 489 occurrences (53.69\%) indicating that many conversations were calm or emotionally steady. This is understandable given that the conversations took place in an environment where participant comfort is paramount. Happy emotions followed with 245 occurrences, showing a considerable amount of positive engagement. In contrast, fear was the least frequent, appearing only 2 times (0.2\%), suggesting minimal instances of fear during the conversations. Surprise was also rare, occurring 21 times, while disgust and sad emotions appeared 110 and 44 times, respectively, indicating moderate levels of negative emotions.
While the dataset, is derived from interactions with a limited number of participants and may not cover the full spectrum of possible human-machine interactions, it offers valuable insights into personalization and emotional engagement within the context of this research. The database details are as follows:

\textit{Textual Data:}
The textual dataset encapsulates a rich array of information derived from the interviews conducted between participants and the robot. This data is structured into CSV and JSON file formats, allowing for easy access and analysis. The key components of the textual dataset are:

\begin{itemize}
    \item \emph{Participant Profile:} containing the participant's information collected from the questionnaire. 
    \item \emph{Chat History:} The dialogues between the robot and the participants are transcribed and included in the dataset. Each exchange is recorded to capture the flow of the conversation and the context within which responses are made.
    \item \emph{Sentiment Analysis:} Utilizing advanced sentiment analysis algorithms, the dataset annotates each conversational exchange with sentiment labels. These labels will provide insights into the emotional undertones of the dialogues, such as positive, negative, or neutral sentiments.
    \item \emph{Emotion Labels:} The dataset documents the emotions exhibited by the participants during the conversation. This label is based on the non-verbal cues, offering a nuanced view of the emotional experienced by the participants throughout the interaction.
    \item \emph{Time Stamps:} Precise time stamps indicating the start and end of each conversation segment. 
\end{itemize}

\textit{Recorded Video and Audio:}
In addition to the textual data, the dataset has a collection of video and audio recordings, corresponding to each individual interview session. The video was recorded using three separate cameras as shown in figure \ref{fig:video_channels}.  These videos and audio provide a dynamic visual and auditory of the interactions, capturing:

\begin{itemize}
    \item \emph{Facial Expressions:} The participants' facial expressions throughout the interview, offer valuable non-verbal cues that complement the sentiment and emotion labels in the textual dataset.
    \item \emph{Voice Modulations:} The tonal variations and nuances in the participants' voices, which are indicative of their emotional states and engagement levels during the conversation.
    \item \emph{Body Language:} The participants' body language, including gestures and posture, which can provide additional context to the verbal communication and emotional expressions.
\end{itemize}

\subsection{Evaluation Criteria}
The evaluation is based on six main criteria: naturalness, engagingness, fluency, relevance, empathy and consistency. Naturalness is a concept that lacks a universally agreed-upon definition \cite{klopfenstein2020study}. Generally, it refers to the ability of conversations to simulate and emulate human conversations \cite{atiyah2019evaluation},  \cite{hung2009towards}. Additionally, naturalness is defined by the dialogue system's capacity to convince users that they are interacting with a human rather than a virtual agent \cite{abushawar2016usefulness}. User engagement is defined as the system's ability to maintain the user's interest and encourage ongoing interaction \cite{yu2016wizard}. While, fluency means that the generated responses are grammatically correct and overall easy to understand which includes the smoothness of the conversation, clear communication, and maintaining a natural conversational flow\cite{tavakoli2016fluency}. The relevance is described as the appropriate and on-topic the system’s responses are with the user’s responses ensuring they contribute meaningfully to the dialogue \cite{bartl2017retrieval}. Finally, empathy and consistency are defined as the ability of the system to demonstrate understanding and sensitivity to the user’s emotions \cite{wang2021transformer} and to generate consistent responses within the conversation context \cite{lin1999consistent} and with the user’s personal information \cite{shi2021novel}, respectively.

\subsection{Automatic Evaluation}
\subsubsection{ PERCY Generating Evaluation}
Due to the absence of ground-truth data for evaluating the generated data, reliance was placed on automatic metrics that do not necessitate reference data, such as perplexity, cosine similarity, MAUVE, and GPT-4 evaluation metrics.

\begin{itemize}
    \item \textit{Metrics}: The perplexity (PPL) is a metric used to assess the fluency and coherence of the generated response \cite{meister2021language}. While, cosine similarity (BERT embeddings) \cite{li2013distance} and MAUVE \cite{pillutla2021mauve} are used to measure the similarity of the generated response with the previous part of the conversation and the distribution of human-like responses, respectively. 
   \item \textit{Results}: The results of PERCY on those metric are as follows: 
   
   \textbf{The perplexity = 24.53}, indicating PERCY's performance in predicting the next word. Lower perplexity is better, showing the model's confidence. A score of 24.53 is moderate, suggesting the model is reasonably adept at generating meaningful responses but occasionally struggles with complex or emotionally nuanced dialogue due to the variability in conversational data.
.
   
   
   \textbf{Cosine similarity = 0.32}, The score of 0.32 indicates moderate alignment with human-generated responses. While the model produces fluent and relevant responses, it may miss deeper emotional or contextual nuances expected in empathetic dialogue.
   
   \textbf{MAUVE = 0.98}, This score suggests that the generated responses are diverse and closely mimic the statistical properties of human responses. The high MAUVE score confirms that the model produces varied and fluent responses that contribute positively to the flow of conversation.

    \item \textit{Metrics}: The GPT-4 model was utilized to assess the overall quality of the generative responses in the study conducted by Wu et al. in 2024 \cite{wu2024gpt}. Similarly, this work employed GPT-4 model evaluation to measure the naturalness, engagement, fluency, relevance, consistency, and empathy of the collected dataset. The use of the GPT model enabled an approximation of the overall quality of the data, mimicking human evaluation.

    \item \textit{Results}: The GPT evaluation is conducted using a Likert scale rating of 5 scores where 5 is the highest and 1 is the lowest. The results indicated very high scores across all measurement criteria, reflecting a high-quality dataset. 
    In terms of naturalness, the average evaluation score was (4.45). While the responses generally sounded natural and facilitated smooth conversation, minor robotic or formal responses had a slight impact on the overall score for some participants. Engagement scored very highly, averaging (4.62), indicating that participants were highly engaged due to relevant follow-up questions, despite minor repetitions affecting engagement slightly. Fluency averaged a score of (4.53), representing a smooth conversation flow and good performance in terms of language clarity and grammar. However, repetitive or overly formal responses slightly disrupted fluency for some users.
    Relevance also scored highly at (4.58), demonstrating that responses were contextually relevant, despite rare off-topic questions. The consistency score of (4.73) indicated a very high level of consistency, suggesting that PERCY was remarkably consistent with users' profiles and the logical flow of the conversation, regardless of some instances where it shifted topics unexpectedly. Finally, the empathy evaluation score was (4.67), showing that the PERCY demonstrated empathy well by acknowledging users' emotions appropriately, even though there were a few cases where more depth could have been applied.
\end{itemize}

\subsection{User Evaluation}
In this study, participants' feedback was utilized to assess the performance of PERCY. Each participant provided feedback through a survey to assess their overall interaction experience. \\
\textit{Conversations Evaluation:}
The feedback yielded promising results, indicating high system performance and a high-quality dataset. The survey results showed that participants had positive feedback for PERCY across all metrics, acknowledging its ability to engage in meaningful, relevant, and engaging conversations. However, areas for improvement were also identified, particularly related to language complexity and topic transitions..

Participants expressed satisfaction with PERCY's conversational abilities, particularly appreciating its capacity to personalize conversations and adapt to various topics. While the robot was perceived as empathetic, engaging, and fluent, there were suggestions for minor improvements in simplifying language and reducing repetitive questions. The naturalness of interactions with the robot was highly rated, with an average of (4.6), with participants noting that conversations felt smooth and human-like. 
PERCY demonstrated a commendable understanding of participants' emotions, as evidenced by its ability to ask relevant follow-up questions and validate their feelings. Although empathy scores were generally high (4.5), a few participants reported that the robot could have expressed empathy more explicitly in some instances. Moreover, participants felt highly engaged throughout the conversations, attributing this to the robot’s interactive and curious nature. As a result, participants evaluated PERCY's engagement with an average of (4.7). 
Indeed, participants found the conversations with PERCY to be fluent and easy to follow with an average of (4.5), whereas minor issues such as misunderstandings or unclear phrasing were noted, which did not significantly disrupt the flow of the conversations. Additionally, the robot's ability to maintain consistent conversation topics and respond appropriately based on participants' inputs was well received. Participants appreciated that the robot adhered to their profile information giving an average score of (4.6) for consistency.
Lastly, the relevance of PERCY's responses was praised with an average score of (4.6), with participants noting its adeptness at staying on topic and maintaining the conversation's focus on the user’s profile information.

The comparison between the user and GPT model evaluations is depicted. The results indicate a high degree of similarity, with the user evaluations having higher scores in naturalness and engagement, while the GPT's evaluations show higher scores in empathy and consistency.

\textit{NASA Task Load Index (NASA TLX):}
The NASA TLX survey provides insight into participants' workload during the robot interview. The scores are rated on a scale from 1 (very low) to 5 (very high) across five dimensions. The average score in each dimension is as follows:

\textit{Mental Demand (2.53):} Participants found the mental demand moderate. This suggests that while the interview required some mental effort in terms of thinking and processing, it was not overly challenging or demanding.
\textit{Physical Demand (1.65):} The physical demand was rated quite low, which is expected in an interview setting where minimal physical effort is required. Participants most likely have not experienced any physical discomfort during the interaction.
\textit{Temporal Demand (2.14):} The score reflects a moderate perception of being rushed or having enough time to answer. While some participants may have felt a little hurried, the majority seemed comfortable with the pace of the conversation.
\textit{Performance (4.06):} This high score indicates that participants successfully expressed their thoughts and feelings during the interview. They were generally satisfied with how they communicated, reflecting confidence in their ability to convey their messages clearly.
\textit{Effort (1.94):} The effort required to perform well during the interview was relatively low. This means that participants did not need significant effort to perform well. This suggests that the interaction felt natural and easy to manage for most individuals.
\textit{Frustration (1.88):} This low score implies that participants did not feel overly annoyed or confused during the interview. The interaction with the robot was smooth and free from significant difficulties.

Overall, The NASA TLX results indicate that participants found the interview mentally engaging but not overly demanding in terms of effort or frustration. The interaction was comfortable, and participants felt they performed well in expressing their thoughts. 

\textit{Negative Attitudes Towards Robots (NARS):}
The NARS aims to understand participants' feelings and comfort levels when interacting with robots. Participants rated their agreement with several statements on a scale of 1 (strongly disagree) to 5 (strongly agree), allowing us to gauge their perspectives on emotional and social interactions with robots. For the statement, “I would feel uneasy if robots really had emotions”, the average score was 2.11. This relatively low score suggests that most participants did not feel significantly uneasy about robots' emotions. The result indicates that participants are generally open to the concept of robots possessing emotional traits, with little discomfort surrounding this possibility.

In contrast, the statement “I would feel relaxed talking with robots” received an average score of 3.71. This score demonstrates a generally positive response, with participants feeling relaxed or neutral about talking with robots. The relatively high score reflects comfort and ease in conversing with robots, implying that most people do not perceive human-robot interactions as intimidating or unsettling under our experimental conditions.
The statement “If robots had emotions, I would be able to make friends with them” yielded an average score of 3.82. This high score shows that participants are largely open to forming social bonds with robots if they are emotionally expressive. Participants appear to believe that the presence of emotions in robots could facilitate more personal and relatable interactions, making it easier to develop a sense of friendship.
Lastly, the statement “I feel comforted being with robots that have emotions” scored an average of 3.76. This score indicates that participants generally feel neutral to slightly positive about feeling comforted by emotionally aware robots. The result suggests that emotional expressions in robots contribute to a sense of comfort and ease during interactions, enhancing the quality of human-robot engagement.
The results reflect a generally positive attitude towards interacting with emotionally aware robots. Participants are open to friendly and comforting interactions with robots expressing emotions. This suggests that participants see value in robots having emotions and believe that this could enhance human-robot interactions, making them more natural and relatable.




\section{Limitations}
The primary constraint of this study lies in the limited number of participants. It is challenging to recruit a substantial number of participants who are willing to be recorded and have videos captured. Furthermore, this research has not received funding from any organization, relying solely on volunteers willing to participate without compensation. However, there are plans to expand this dataset to encompass a broader and more diverse range of individuals.
Another limitation is that this dataset is based on open-domain conversations, making it difficult to cover all open-domain topics. Notably, one of our potential objectives is to create a domain-specific dataset, which would aid in collecting more precise personal data. This, in turn, could be utilized to develop a potentially new dataset and system. 

\section{Conclusion}
This research contributes to the development of emotionally intelligent and personalized conversational systems. By creating a novel multimodal dataset (MERCI) derived from real human-robot interactions, this study aims to advance the field of human-robot communication. The dataset combines personal information with emotional cues to enable the training of conversational models to provide empathetic, relevant, and engaging responses. Despite challenges such as a limited participant pool and the open-domain nature of the dataset, the study demonstrates promising results in enhancing personalization and empathy in dialogue systems, paving the way for future improvements in personalization and emotionally aware human-robot interactions.

\bibliographystyle{IEEEtran}
\bibliography{reference}

\newpage 
\appendices

\end{document}